\documentclass{article}

\usepackage{PRIMEarxiv}

\usepackage[utf8]{inputenc} 
\usepackage[T1]{fontenc}    
\usepackage{hyperref}       
\usepackage{url}            
\usepackage{booktabs}       
\usepackage{amsfonts}       
\usepackage{nicefrac}       
\usepackage{microtype}      
\usepackage{lipsum}
\usepackage{fancyhdr}       
\usepackage{graphicx}       
\graphicspath{{media/}}     

\title{Mapping Parallel Matrix Multiplication in GotoBLAS2\\ 
       to the AMD Versal ACAP for Deep Learning}

\author{
  Jie Lei, Enrique S. Quintana-Ortí \\
  Universitat Politècnica de València \\
  Valencia \\
  Spain \\
  \texttt{\{jlei, quintana\}@disca.upv.es} \\
}

\usepackage{pgfplots}
\pgfplotsset{width=10cm,compat=1.14}

\usepackage{listings}

\usepackage{amsmath}
\usepackage{amsfonts}
\usepackage{graphicx}
\usepackage{textcomp}
\usepackage{xcolor}
\usepackage{listings}
\usepackage{hyperref}
\usepackage{url}
\usepackage{xspace}
\usepackage{enumitem}

\usepackage[misc,geometry]{ifsym}
\usepackage[export]{adjustbox}
\usepackage{listings}
\usepackage{cleveref}

\usepackage{xcolor}
\definecolor{LightGray}{gray}{0.9}

\def\BibTeX{{\rm B\kern-.05em{\sc i\kern-.025em b}\kern-.08em
    T\kern-.1667em\lower.7ex\hbox{E}\kern-.125emX}}

\newcommand{\gemm}{\textsc{gemm}\xspace}

\newcommand{\pe}{\mathrel{+\!\!=}}

\definecolor{gray98}{rgb}{0.93,0.93,0.93}
\definecolor{gray20}{rgb}{0.20,0.20,0.20}
\definecolor{gray25}{rgb}{0.25,0.25,0.25}
\definecolor{gray16}{rgb}{0.161,0.161,0.161}
\definecolor{gray60}{rgb}{0.6,0.6,0.6}
\definecolor{gray30}{rgb}{0.3,0.3,0.3}
\definecolor{bgray}{RGB}{248, 248, 248}
\definecolor{amgreen}{RGB}{40, 144, 40}
\definecolor{myblue}{RGB}{0, 40, 255}
\definecolor{amred}{RGB}{228,26,28}
\definecolor{amethyst}{rgb}{0.6, 0.4, 0.8}
\definecolor{mymauve}{rgb}{0.58,0,0.82}
\definecolor{LightGray}{gray}{0.9}

\usepackage[hyperfirst=false]{glossaries}
\newacronym{ai}{AI}{Artificial Intelligence}
\newacronym{ml}{ML}{Machine Learning}
\newacronym{hci}{HCI}{Human-Computer Interaction}
\newacronym{exp}{EXP}{Example}
\newacronym{ps}{PS}{Processing System}
\newacronym{aie}{AIE}{AI Engine}
\newacronym{fpga}{FPGA}{Field Programmable Gate Array}
\newacronym{pl}{PL}{Programmable Logic}
\newacronym{lut}{LUT}{Lookup Table}
\newacronym{mac}{MAC}{Multiplication and Accumulation}
\newacronym{ddr}{DDR}{Dynamic Random-Access Memory}
\newacronym{gmio}{GMIO}{Global Memory Input Output}
\newacronym{kb}{KB}{Kilobyte}
\newacronym{mb}{MB}{Megabyte}
\newacronym{axi}{AXI}{Advanced eXtensible Interface}
\newacronym{noc}{NOC}{Network on Chip}

\usepackage[normalem]{ulem}

\begin{document}
\maketitle

\newcommand{\mus}{$\mu\text{s}$\xspace}
\newcommand{\macpc}{$MACs/cycle$\xspace}

%

\begin{abstract}
This paper investigates the design of parallel general matrix multiplication (\gemm) for a Versal Adaptive Compute Accelerated Platform (ACAP) equipped with a VC1902 system-on-chip and multiple Artificial Intelligence Engines (AIEs). 
Our efforts aim to port standard optimization techniques applied in the
high-performance realization of \gemm on CPUs 
to the Versal ACAP. In particular, 
1) we address the flexible exploitation of
the Versal ACAP's multi-level memory hierarchy; 
2) we delve into the efficient use of the vector units in the AIE tiles,
proposing an architecture-specific micro-kernel 
for mixed precision arithmetic
to address the strong demand for adaptive-precision inference in deep learning;  and
3) we introduce a parallel design for \gemm that spans multiple AIE tiles, enhancing the computational throughput. 
We conduct experimental profiling,
with up to 32 AI Engines, 
that demonstrates the high parallel scalability of the solution.

\end{abstract}

\section{Introduction}

Over the past three decades, the deceleration of Moore's Law and Dennard scaling has hindered the advancement of single-core 
computer architecture performance~\cite{4785534,Eec17}. In the mid-2000s, the adoption of multicore processors emerged as a response to this challenge first; 
and this was followed by domain-specific accelerators, 
such as NVIDIA's tensor cores, Google's MXU in the Tensor Processing Units (TPUs), 
Intel's AMX, etc.~\cite{silvano2023survey}. 
Recognizing the benefits of specialized hardware from the perspective of both performance and power, AMD/Xilinx introduced the Versal Adaptive Compute Accelerated Platform (ACAP) in 2019. This architecture integrates high-performance SIMD (single instruction, multiple data) processors, sophisticated input/output capabilities, and integrated memory controllers, accommodating a diverse range of workloads in general, and deep learning (DL) in particular~\cite{8875639,9220682}.

The paper is rooted in our previous work, where we developed an initial design focused on a single-core architecture utilizing INT16 precision arithmetic~\cite{10136983}.
We further investigate the mapping of the standard formulation of the parallel general matrix multiplication (\gemm) for multicore conventional processors, as exemplified in GotoBLAS2~\cite{Goto:2008:AHP,BLIS1,BLIS2}, onto a
Versal ACAP is equipped with multiple artificial intelligence engines (AIEs). We are particularly interested in this computational kernel because \gemm serves as the cornerstone for the software packages upon which 
a myriad of scientific and engineering codes are built. 
Moreover, DL training and inference with well-known convolutional neural networks (CNNs), as well as modern transformer encoders, cast a significant portion of their arithmetic cost in terms of this computational kernel~\cite{Che06,CHITTYVENKATA2023102990,kim2023stack}.

In addressing the efficient implementation of \gemm in the Versal ACAP, our work makes the following contributions:
\begin{itemize}
\item \textbf{Single core/tile and memory hierarchy}: We provide a brief elaboration of the techniques employed in \\ high-performance realizations of \gemm (e.g., in GotoBLAS2, OpenBLAS, BLIS, AMD AOCL, ARM PL, etc.). Furthermore,
we discuss how these techniques exploit the SIMD units and cache memory, and then explain how to adapt and leverage these ideas for the multi-level memory of the Versal ACAP.
\item \textbf{Low-Precision Inference}: To address the demand for low-precision inference in DL, we propose an architecture-specific micro-kernel designed to operate with mixed precision in the Versal AIE SIMD units.
This is a fundamental component in high-performance \gemm.
\item \textbf{Parallel Design}: We introduce a parallel design for \gemm that runs on multiple AIE tiles. We then conduct a theoretical analysis and an experimental performance profiling of our scheme involving up to 32 AI Engines.
\end{itemize}

The subsequent sections of this paper follow a structured approach. 
In~\Cref{sec:gemm}, we introduce the foundational concepts of high-performance \gemm and provide a general overview of the Versal ACAP. 
\Cref{sec:acap_intro} briefly presents the critical features of the Versal ACAP. Moving on 
to~\Cref{sec:mainDesign}, we detail our design of \gemm, focusing on the memory mapping for the matrix operands. 
This section also elucidates the use of AI Engines for micro-kernel 
execution, discussing the distribution of the \gemm iteration space
across various SIMD AIE tiles. Finally, \Cref{sec:profiling} presents a comprehensive performance analysis of the multiple SIMD design. Within this section, we identify communication bottlenecks and outline potential strategies for further optimization.

\section{High Performance GEMM on Conventional Architectures} \label{sec:gemm}

\begin{figure*}[t]
    \centering
    \begin{tabular}{ccc}
        \begin{minipage}[t]{0.35\textwidth}
            \begin{tabular}{l}
                \footnotesize
\begin{lstlisting}[language=C,alsoletter={.},deletekeywords={.sum},morekeywords={}]
// Baseline algorithm for GEMM
for (jc=0; jc<n; jc+=nc) // Loop L1
 for (pc=0; pc<k; pc+=kc){    // L2
  // Pack B
  Bc:=B(pc:pc+kc-1, jc:jc+nc-1);
  for (ic=0; ic<m; ic+=mc){   // L3
   // Pack A
   Ac:=A(ic:ic+mc-1, pc:pc+kc-1);
   for (jr=0; jr<nc; jr+=nr)  // L4
     for(ir=0; ir<mc; ir+=mr) // L5
      // Micro-kernel
      C(ic+ir:ic+ir+mr-1, jc+jr:jc+jr+nr-1)
        += Ac(ir:ir+mr-1, 0:kc-1)
        *  Bc(0:kc-1, jr:jr+nr-1);
 }}
\end{lstlisting}
                ~\\
                ~\\
                ~\\
                \footnotesize
\begin{lstlisting}[language=C,alsoletter={.},deletekeywords={.sum},morekeywords={}]
// Micro-kernel
for (pr=0; pr<kc; pr++) // Loop L6
  C(ic+ir:ic+ir+mr-1, jc+jr:jc+jr+nr-1)
    += Ac(ir:ir+mr-1,pr) 
    *  Bc(pr,jr:jr+nr-1);
\end{lstlisting}
                ~\\
                ~\\
                ~\\
                \includegraphics[width=0.9\textwidth]{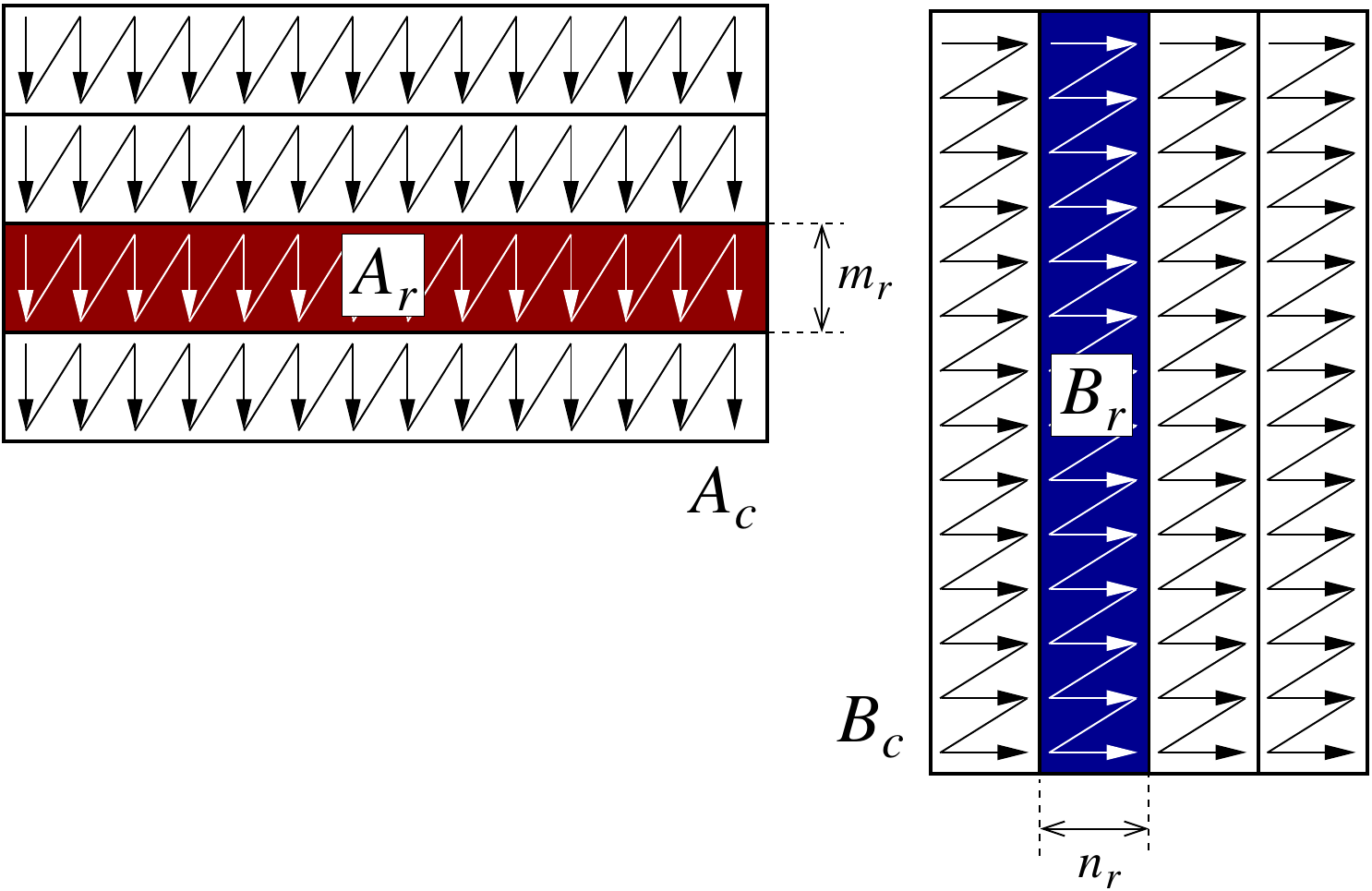}
            \end{tabular}
        \end{minipage}
        & \hspace*{10ex} &
        \begin{minipage}[t]{0.5\textwidth}
            \begin{tabular}{c}
                \includegraphics[width=0.8\textwidth]{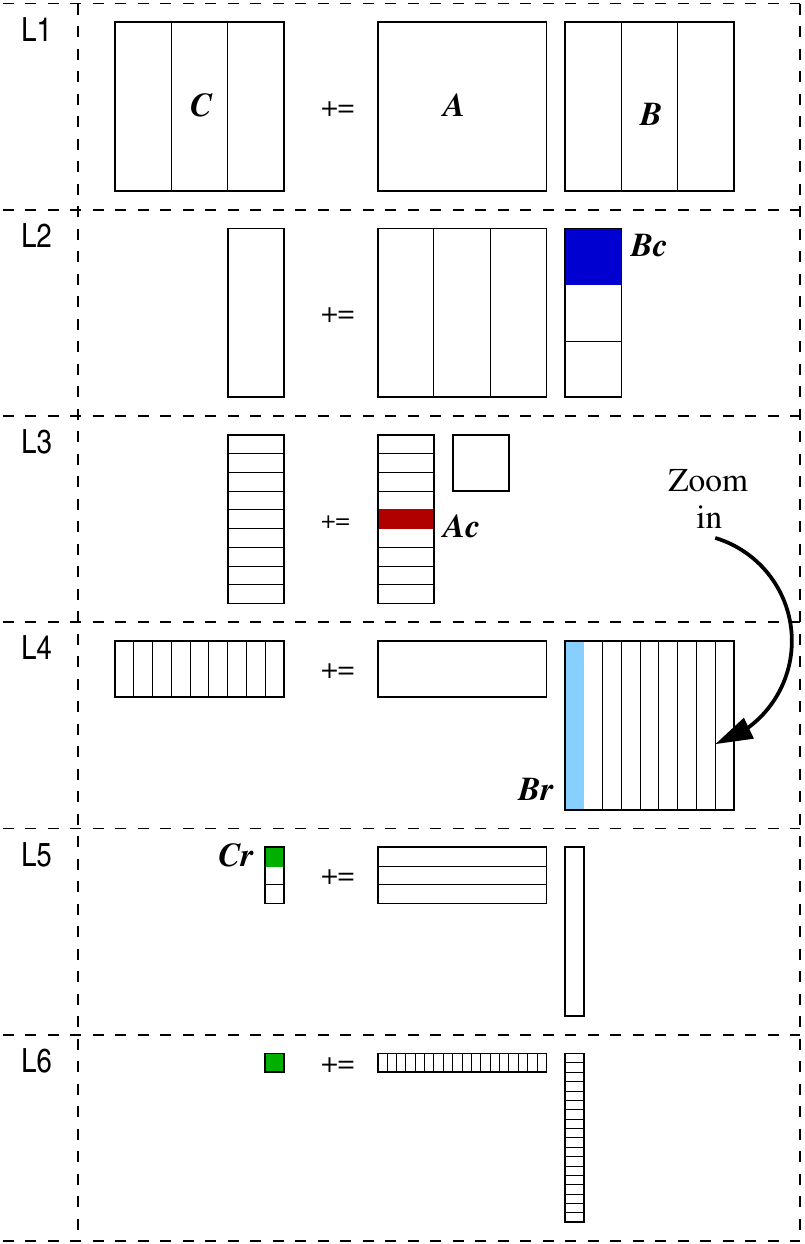} \\
                ~\\ 
                \includegraphics[width=0.8\textwidth]{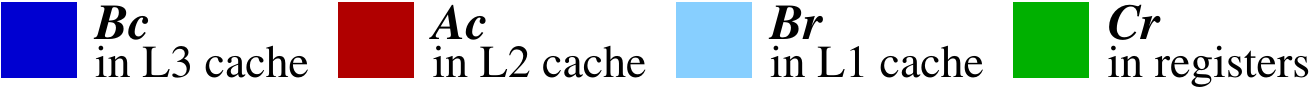}
            \end{tabular}
        \end{minipage}
    \end{tabular}
    \caption{Baseline high performance algorithm for \gemm. 
    Top-Left: Blocked algorithm; Middle-Left: Micro-kernel; Bottom-Left: Packing of input matrix operands.
    Right: Data transfers across the memory hierarchy.}
    \label{fig:baseline_GEMM}
        \vspace*{-2ex}
\end{figure*}

Consider the \gemm $C \pe AB$, where the matrices $A$, $B$, and $C$ have dimensions $m \times k$, $k \times n$, and $m \times n$, respectively. Modern instances of this key computational kernel, integrated into high-performance libraries for scientific computing and deep learning such as OpenBLAS, BLIS, AMD AOCL, ARM PL, and Intel oneAPI, adhere to the ideas introduced by GotoBLAS2~\cite{Goto:2008:AHP}. 
In particular, they formulate the kernel 
as five nested loops plus two packing procedures and 
a \textit{micro-kernel};
see the algorithms in the top-left and middle-left of \Cref{fig:baseline_GEMM}. 
Take note there of the loops labeled as \textsf{L1},\textsf{L2},\ldots,\textsf{L5} in the 
\gemm algorithm. 
 
In this baseline algorithm for \gemm, an appropriate selection of the strides (also called cache configuration parameters, or CCPs) for 
the three outermost loops --$n_c, k_c, m_c$--, along with a careful copy of certain blocks of the input matrix operands
into the buffers $A_c, B_c$, respectively of dimensions $m_c \times k_c,k_c \times n_c$, (see 
Figure~\ref{fig:baseline_GEMM}, bottom-left,) significantly reduce the volume cache misses~\cite{BLIS1,BLIS4}. 
Specifically, these three configurable parameters 
should be tailored to the cache memory of the target architecture to map certain blocks of
the matrix operands to specific levels of the cache hierarchy; 
see Figure~\ref{fig:baseline_GEMM}, right. 
For simplicity, we shall assume that $m, n, k$ are integer multiples of $m_c, n_c, k_c$ respectively.

In addition, the micro-kernel implements an extra loop (labeled as \textsf{L6} in \Cref{fig:baseline_GEMM}, middle-left) that updates a small block of matrix $C$ of dimension $m_r \times n_r$, known as the micro-tile $C_r$, 
with the product of two sub-blocks from $A_c,B_c$, of dimensions
$m_r \times k_c, k_c \times n_r$, and which are referred to as
the micro-panels $A_r,B_r$; see the high-lighted sub-blocks in~\Cref{fig:baseline_GEMM}, bottom-left. 
This is achieved by performing a sequence of $k_c$ rank-1 transformations in the micro-kernel loop body, 
each involving the $m_r$ elements in a column of $A_r$ and the $n_r$ elements in a row of $B_r$. 
In contemporary processors with SIMD units, the micro-kernel dimensions $m_r \times n_r$ are chosen to enable vectorization within the micro-kernel loop \textsf{L6} while avoiding register spilling. The packing facilitates a given arrangement of the data in memory that allows the micro-kernel to load the micro-panel entries with unit stride utilizing SIMD instructions.

\section{Architecture of the Versal ACAP}\label{sec:acap_intro} 

\begin{figure}[htbp]
    \centering
    \includegraphics[width=0.6\columnwidth]{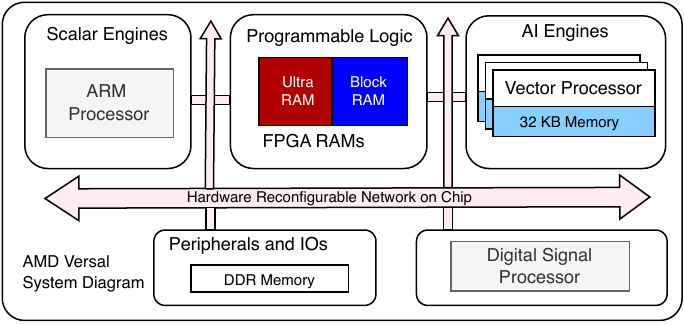}
    \caption{Block diagram of the Versal AI Core.
    }
    \label{fig:acap}
        \vspace*{-3ex}
\end{figure}

The AMD Versal VC1902, with the organization depicted in Figure 
\ref{fig:acap}, is a heterogeneous system-on-chip (SoC) offering high arithmetic throughput combined with a flexible memory architecture. 
The design comprises three computing systems: 1) an ARM Cortex-A72 processor; 2) an
FPGA (Filed Programmable Gate Array) with 899,840 LUTs (Lookup Tables); and 3) an array of 400 high-compute throughput vector AIEs. 
The ARM processor is in charge of orchestrating the data transfers between the modules and can also perform 
general-purpose computing tasks, such as data manipulation. 
The FPGA can be used for low-latency compute acceleration and memory storage to assist in high-throughput operations.  
Lastly, the AIEs comprise SIMD arithmetic units 
that support mixed-precision computing and deliver up
to 128 (8-bit integer) 
GigaMAC (Multiply-and-Accumulate) operations per second at their peak.

\section{Customizing GEMM for the Versal ACAP}\label{sec:mainDesign}

This section describes our efforts to
reformulate the conventional high-performance realization of \gemm in modern libraries 
to the architecture of the Versal ACAP.
In particular, this was achieved by tailoring three main
factors, to be described in detail in the following subsections: 
\begin{enumerate}
\item Map the matrix operands to the proper levels of the Versal ACAP memory hierarchy;
\item Exploit the SIMD units in the artificial intelligence engine (AIE) tiles  from within the micro-kernel; and
\item Select the appropriate \gemm loop to parallelize, distributing the corresponding iteration space 
      evenly across the grid of AIE tiles.
\end{enumerate}

\subsection{Distributing the operands across the memory hierarchy}

\begin{table}[htb]
    \centering
    \caption{
    \textcolor{black}{Multi-level memory hierarchy in the Versal VCK190.}
    }
    \vspace*{-2ex}
    \begin{tabular}{lrll}
    \toprule
    Memories  & Capacity & Operands & Cache\\
    \midrule
    AIE tile vector registers    &  2 KB  & $C_r$  & Registers \\ 
    AIE tile local memory   &  32 KB & $B_r$  & L1\\
    FPGA Ultra RAM          &  16.27 MB & $A_c, A_r$& L2\\
    FPGA Block RAM          &  4.25 MB & $B_c$      & L3\\
    DDR4 global memory    &  2 GB  & $A, B, C$      & RAM        \\
    \bottomrule
    \end{tabular}
    \label{tab:multi_level_cache}
    \vspace*{-2ex}
\end{table}


Table~\ref{tab:multi_level_cache} displays the four memory levels that we target in our customization 
of \gemm for the Versal ACAP, ordering them from top/fastest to bottom/slowest.
In general, the dimensions of the \gemm operands are large and exceed the storage capacity of the fastest
yet smallest levels in the
the memory hierarchy of the computing device. The natural strategy thus involves decomposing the problem into 
several more manageable 
subtasks, each operating with a part of the matrix operands, and keep their data close 
to the arithmetic units. 
In summary, the goal is to distribute the different matrix operands of \gemm to the various memory layers, depending on their
capacities and bandwidths, to augment data reusability and reduce communication overhead
(data transfers)
during the execution of the \gemm algorithm.

In detail, the blocking strategy applied by the three outermost loops of the \gemm algorithm 
in the GotoBLAS2 scheme
(\textsf{L1},
 \textsf{L2},
 \textsf{L3})
cast the calculation into a collection of smaller matrix multiplications.
Specifically, each of these subtasks 
multiplies the contents of the buffers $A_c$ and $B_c$, to update a certain
block of $C$, which we denote as $C_c$. 
(Note the difference between $A_c,B_c$, which are actual copies of certain blocks
of the input matrix operands; and $C_c$, which is an artifact introduced to ease
the notation.)
Internally, the next two loops
(\textsf{L4},
 \textsf{L5})
decompose the operation $C_c \pe A_c\, B_c$ into a series of
even smaller subtasks, each performing a product of the type $C_r \pe A_r \, B_r$,
corresponding to a micro-kernel. 


Due to their size, the input/output matrix operands $A,B,C$ are kept in the 
large yet slow global memory.
From bottom to top in the memory hierarchy, 
the buffer $B_c$ is maintained in the low throughput FPGA Block RAM.
In a conventional architecture with a hardware-assisted cache, this is achieved transparently to the programmer/user,
due to the particular ordering of the three outermost loops, provided 
an appropriate selection of the cache configuration parameters $k_c,n_c$ is made. 
In contrast, in the Versal ACAP this is enforced from within the packing routine for $B_c$, by explicitly copying
the data block from $B$ in the global memory 
into the memory space for $B_c$ in the FPGA Block RAM.
%

The GotoBLAS2 scheme dictates that the data for the buffer $A_c$ is copied
and packed into the L2 cache memory. 
In our case, the benefits of quickly streaming the data from the FPGA memory, 
instead of accessing those from the global memory,
guided us to pack this buffer into the high throughput FPGA Ultra RAM.
This is achieved via the packing routine for $A_c$.
Note that having both $A_c$ and $B_c$ in the ``same'' memory level requires splitting the capacity of the FPGA between the
two operands. Nonetheless, we avoid misses due to conflicts because these data are copied/accessed manually 
and logically separated into the FGPA Ultra and Block RAMs.

The GotoBLAS2 scheme emphasizes the exploitation 
of data reusability for higher compute throughput across all levels of the memory hierarchy, including the L1 cache.
In addition to the packing routines, in our solution
this is achieved by utilizing the local memory of the AIE tile 
to store the micro-panel $B_r$.
Note that loop \textsf{L5} utilizes the 
same instance of $B_r$ in multiple operations, each with a different instance of $A_r$. This 
implies that the cost of transferring $B_r$ from the FPGA memory to the local memory can be amortized
over multiple executions of the micro-kernel (one per iteration of loop \textsf{L5}).
Hence, storing this operand in a fast memory is beneficial, to safeguard the computation efficiency.

\begin{figure}[htbp]
    \centering
    \includegraphics[width=0.5\columnwidth]{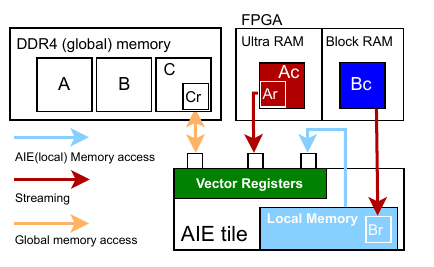}
    \caption{Mapping of \gemm operands to the Versal ACAP memory hierarchy.}
    \label{fig:aie_operand_map_simple}
\end{figure}

Finally, as we will see in the following subsection, at the end of its execution, the micro-kernel copies
the contents of a micro-tile $C_r$ from the main memory to the AIE tile registers, updates its entries, and stores that 
piece of data back into the main memory.

\Cref{fig:aie_operand_map_simple} offers a graphical summary of the strategy that we follow to map the matrix
operands for \gemm to the five levels of the Versal ACAP memory hierarchy:
main memory, FPGA Block/Utra RAM, local memory, and tile registers.
We emphasize that, unlike a conventional computing system, the Versal ACAP does not have a cache controller 
to orchestrate the data movements between the different levels transparently to the programmer. 
Hence, the actual implementation of our \gemm algorithm 
requires explicitly performing the data transfers between the distinct memory
levels, including the AIE tile vector registers. This is done from inside the packing routines as well as the micro-kernel.
Furthermore, we exploit the platform's family of communication 
protocols: the global input-output interface, the streaming interface, and direct memory copies.
This allows us to accommodate the algorithm to the multi-level memory system in this platform,
mimicking the GotoBLAS2 scheme despite lacking a hardware-assisted cache controller.

\subsection{SIMD Micro-kernel for the AIE tile}\label{sec:microKernel}

In line with the prevailing practice of utilizing low-precision arithmetic for DL inference, 
we choose UINT8 as the baseline data type for our parallel implementation of \gemm targeting the Versal ACAP. 
Considering the number of 
accumulator registers in the AIE tile and their capacity, we set the dimensions of the micro-tile $C_r$ that
is updated inside loop \textsf{L6} of the micro-kernel to $m_r \times n_r = 8 \times 8$. 
This fully utilizes the accumulator registers,
matches the functionality of the \texttt{mac16()} intrinsic, and provides
the potential to deliver 128 MACs/cycle for the UINT8 data type.

\begin{figure}[tbh!]
\begin{minipage}[t]{\columnwidth}
\begin{lstlisting}[language=C++,alsoletter={.},deletekeywords={.sum},morekeywords={v32uint8,v64uint8,v16acc48,input_window_int8,int16,output_window_int16,input_stream_int8,__restrict},numbersep=6pt]
void micro_kernel( input_window_int8 * __restrict DDR_IN,
                   input_stream_int8 * __restrict PL_IN,
                   int16 *Br,
                   output_window_int16 *__restrict out){


  // Vectors for ar, br; Accumulators for C
  v64uint8 ar0, ar1;
  v32uint8 br;
  v16acc48 Cacc0, Cacc1, Cacc2, Cacc3;

  // Parameters for mac16() intrinsics
  unsigned int xoffsets   = 0x11101110;
  unsigned int xsquare    = 0x3120;
  unsigned int zoffsets_0 = 0x04040000;
  unsigned int zoffsets_1 = 0x0C0C0808;
  unsigned int zsquare    = 0x1010;

  for (unsigned int i=0; i<kc; i+=16){ // Loop L6
    // Unrolled x16 to overlap transfers and computations
    chess_prepare_for_pipelining;
    chess_loop_range(kc); 

    // Read 64x2 UINT8 elements of Ar into ar0, ar1 
    ar0 = readincr_v64(PL_IN);
    ar1 = readincr_v64(PL_IN);

    // Repeat 4x: Read 32 UINT8 elements of Br and compute
    br = *(v32uint8*) Br[i];
    Cacc0 = mac16(Cacc0, ar0, 0, xoffsets, 16, xsquare, 
                  br, 0, zoffsets_0, 2, zsquare);
    Cacc1 = mac16(Cacc1, ar0, 0, xoffsets, 16, xsquare, 
                  br, 0, zoffsets_1, 2, zsquare); 
    
    br = *(v32uint8*) Br[i+8];
    Cacc2 = mac16(Cacc2, ar0, 0, xoffsets, 16, xsquare, 
                  br, 0, zoffsets_0, 2, zsquare);
    Cacc3 = mac16(Cacc3, ar0, 0, xoffsets, 16, xsquare, 
                  br, 0, zoffsets_1, 2, zsquare); 

    br = *(v32uint8*) Br[i+16];
    Cacc0 = mac16(Cacc0, ar1, 0, xoffsets, 16, xsquare, 
                  br, 0, zoffsets_0, 2, zsquare);
    Cacc1 = mac16(Cacc1, ar1, 0, xoffsets, 16, xsquare, 
                  br, 0, zoffsets_1, 2, zsquare); 
    
    br = *(v32uint8*) Br[i+24];
    Cacc2 = mac16(Cacc2, ar1, 0, xoffsets, 16, xsquare, 
                  br, 0, zoffsets_0, 2, zsquare);
    Cacc3 = mac16(Cacc3, ar1, 0, xoffsets, 16, xsquare, 
                  br, 0, zoffsets_1, 2, zsquare); 
  }
  // Read Cr from global memory
  v64uint8 C = window_readincr_v64(DDR_IN);

  // Convert result, add it to Cr, and write back to memory
  C = concat(Cacc0, Cacc1, Cacc2, Cacc3); 
  window_writeincr(out,C);
}
\end{lstlisting} 
\end{minipage}
\caption{ \textcolor{black}{Simplified version of the $8 \times 8$, UINT8 micro-kernel for the AIE tile.} }
\label{alg:microK_AIE_2024}
    \vspace*{-3ex}
\end{figure}

Our implementation of the \gemm micro-kernel tailored to the AIE tile is displayed in ~\Cref{alg:microK_AIE_2024}. 
After a series of declarations and initializations at the beginning of the routine, the micro-kernel loop 
\textsf{L6} (Line~19) iterates across the $k_c$ dimension of the micro-panels $A_r, B_r$ with an unrolling factor of 16. 
At each iteration, the micro-kernel multiplies the elements in 16 columns of the micro-panel $A_r$ 
(loading 64 elements to \texttt{ar0} and 64 more to \texttt{ar1}; lines 25--26) with the elements in 16 rows of 
the micro-panel $B_r$ 
(loading four times 32 elements to \texttt{br}; lines 29, 35, 41, and 47). 
The intermediate results are accumulated in four 
accumulators, using the AIE tile intrinsic function \texttt{mac16()}, which is called eight times in lines 30--51.

The function \texttt{mac16()} performs 128 UINT8 MAC operations in a single cycle (that is, 256 UINT8 operations), 
involving two vectors, respectively with 64 and 32 elements of $A$ and $B$. (Each \texttt{mac16()} only uses
32 of the vector with the elements from  $A$.) Since the AIE tile intrinsics cannot use \texttt{v8uint8}, we thus combine two column vectors of the micro-tile $C_r$ into a single \texttt{v16uint8} variable to carry out the MAC operations.

During each iteration of the loop, 256 elements are fetched from the memory levels near the tile arithmetic units (128 from $A_r$, in the FPGA Ultra RAM; and 128 from $B_r$, in the local memory). 
These elements compute 2,048 UINT8 arithmetic operations (i.e., 1,024 MAC). 
The purpose is to amortize the cost of the memory transfers with a large
amount of arithmetic.  
The extensive usage of the accumulator and vector registers, at 100\% and 75\% respectively, 
together with the careful selection of compiler optimization options, 
enables the concurrent execution of MAC procedures and data transfers.

Upon completion of the loop, the code accumulates the partial result to the contents of $C$ in the global memory. To achieve this, the micro-kernel loads the appropriate $8 \times 8$ micro-tile $C_r$ from the global memory (line 54); updates its contents; and stores the results back in global memory (line 58). The cost of transferring these data between the main memory and
the AIE tile vector registers can be amortized provided $k_c$ is large.
\subsection{Setting the cache configuration parameters}

Returning to the adaptation of the \gemm algorithm to the multi-level memory hierarchy of the Versal ACAP, 
we remind that the blocking strategy aims to re-utilize the data that 
is close to the arithmetic units by implementing the packing schemes shown in the right of \Cref{fig:baseline_GEMM}. 
In consequence,  the CCPs $m_c, n_c, k_c$ must be selected with caution 
since they have a direct impact on the re-utilization factor,
memory consumption, and cache hit ratio for both the local memory and FPGA RAMs.
We next discuss how we chose the CCPs for the Versal ACAP. 

We begin by examining the small local memory intended to hold the micro-panel $B_r$,
of dimension $k_c \times n_r$. 
Now, since the design of the micro-kernel ``hardwires'' $n_r=8$ (and $m_r=8$),
the largest dimension for $k_c$ is in principle constrained 
by the capacity of the local memory (32~KB) divided by the product of $n_r$ and the
size of the data type (1 byte per element for UINT8). 
In practice, we ascertain an upper limit of
3,750 for $k_c$, sparing about 2.5~KB 
for other data that also has to reside in the local memory.

Our next goal is to identify the dimensions of the buffers 
$A_c and B_c$ are stored in the FPGA RAMs. 
The Versal ACAP targeted in this work 
has two types of FPGA memory: Ultra RAM and Block RAM, 
with capacities of 16.3 MB and 4.25 MB, respectively. The Ultra RAM is a high-speed storage medium used in our design to hold the
buffer $A_c$, dimensions $m_c \times k_c$. 
Given that we have already set the upper limit for $k_c$ to 3,750, 
we can deduce that the maximum value for $m_c$ is about 4,500 if we exhaust 
the FPGA Ultra RAM. 
Due to the lower frequency of updates compared to matrix $A_c$, we opt to store matrix $B_c$ in the FPGA Block RAM. 
The dimensions of the matrix are determined by the values of $k_c$ and $n_c$. 
Following an analogous procedure applied to the previous 
calculation, the maximum value for $n_c$ is derived as 1,200.

\subsection{Parallelization of GEMM for the AIE tile grid} \label{sec:parallelGEMM}
 
The previous parts of this section
presented a tailored design for the Versal ACAP, based on the reference algorithm
for \gemm in~\Cref{sec:gemm},
that pursues a sequential implementation on a single AIE tile.
We next extend our design to a parallel \gemm method that harnesses a fraction of the
400 AIE tiles in the Versal ACAP.  

High-performance implementations of \gemm on a multi-core platform usually exploit loop parallelism considering the underlying cache hierarchy~\cite{BLIS2}. 
Concretely, parallelizing loop \texttt{L1} of the baseline algorithm
for \gemm is the preferred option on multi-socket platforms.
In contrast, loop \texttt{L3} is best suited for systems with 
private L1/L2 cache levels and shared L3 cache.
Moreover, parallelizing either loop \texttt{L4} or loop \texttt{L5}
is a fair option for platforms with private L1 cache but shared 
L2/L3 levels.
The latter matches the memory hierarchy of the Versal ACAP, with a local memory 
per AIE tile but shared FPGA Ultra and Block RAMs.
To increase the granularity of the parallelization, and
assuming that $n_c/n_r$ is significant, we thus choose to target
loop \texttt{L4}. To close this discussion, we note that
parallelizing loops \texttt{L2}, \texttt{L6} should be avoided due to potential race conditions.  

\begin{figure*}[htbp]
    \centering
    \includegraphics[width=1\textwidth]{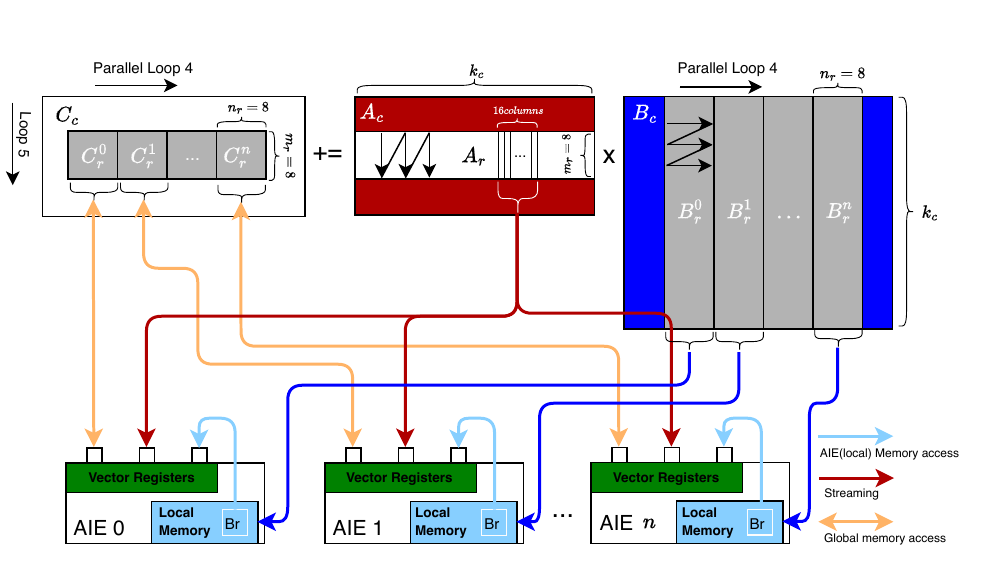}
    \vspace*{-2ex}
    \caption{Data transfer scheme for multiple AIE tiles. Each engine 
    receives a different micro-tile $C_r$ and micro-panel
    $B_r$, but they all share the same micro-panel $A_r$. }
    \label{fig:multi_aie_design}
\end{figure*}

The algorithm showcased in \Cref{alg:unroll_gemm_for_AIE} exemplifies the code 
of our parallel \gemm for the Versal ACAP. 
The parallelization strategy preserves the overall design of loops 
\textsf{L1}--\textsf{L3} in the baseline algorithm for \gemm, but distributes
the iteration space for Loop  \textsf{L4} across {\small \texttt{NUM\_AIEs}} AIE tiles.
The parallelization in the Versal ACAP is graphically illustrated in \Cref{fig:multi_aie_design}.
Each AIE tile copies a distinct micro-panel $B_r$ from the buffer $B_c$ into its local memory.
Then, during the execution of the micro-kernel, each AIE tile accesses the data for
the $B$ operand from its micro-panel $B_r$ in its local memory, while
retrieving the data for the $A$ operand from the same micro-panel $A_r$ 
in the FPGA Ultra RAM shared by all AIE tiles.

At the end of the micro-kernel execution, each AIE tile consolidates the partial updates accumulated to a distinct
micro-tile $C_r$ on the output matrix $C$ stored in the DDR memory.

\begin{figure}[tbh!]
\begin{minipage}[t]{0.88\columnwidth}
\begin{lstlisting}[language=C++,alsoletter={.},deletekeywords={.sum},morekeywords={v32int16,v16int16,v16acc48,input_window_int16,int16,output_window_int16},numbersep=6pt]
for (jc=0; jc<n; jc+=nc) // Loop L1
 for (pc=0; pc<k; pc+=kc){    // L2
  // Pack B
  Bc:=B(pc:pc+kc-1, jc:jc+nc-1);
  for (ic=0; ic<m; ic+=mc){   // L3
   // Pack A
   Ac:=A(ic:ic+mc-1, pc:pc+kc-1);
   // Parallel loop L4 
   for (jr=0; jr<nc; jr+=nr*NUM_AIEs){  
    // Each AI engine copies a distinct 
    // micro-panel Br to its local memory
    Br:=Bc(:, jr:jr+nr-1);
    for(ir=0; ir<mc; ir+=mr){ // L5
     // Micro-kernel, same as in with sequential case
     // All AI engines read data from the same
     // micro-panel Ar,
     // but its own micro-panel Br
     C(ic+ir:ic+ir+mr-1, jc+jr:jc+jr+nr-1)
       += Ac(ir:ir+mr-1, 0:kc-1)
       *  Br(0:kc-1,:);
 }}}}
\end{lstlisting} 
\end{minipage}
\caption{ \textcolor{black}{Simplified parallel implementation of \gemm for the Versal ACAP.} }
\label{alg:unroll_gemm_for_AIE}
    \vspace*{-2ex}
\end{figure}

\subsection{Communication protocols} 

Our design utilizes several communication protocols to reduce the cost of the data transfers, 
recognizing that this type of overhead may be a constraining factor even for compute-intensive algorithms such as \gemm.
The \gemm design utilizes the FPGA RAMs to store the buffers $A_c$ and $B_c$, the local memory of the AIE tiles for
the micro-panels $B_r$, and the vector registers for the micro-tiles $C_r$, 
taking advantage of the re-usability factors intrinsic to the algorithm. 
Specifically, 
the same buffer $B_c$ is accessed once per iteration of loop \textsf{L3} (that is, $m/m_c$ times);
the same buffer $A_c$ is accessed once per iteration of loop \textsf{L4} ($n_c/n_r$ times); 
the same micro-panel $B_r$ is accessed once per iteration of loop \textsf{L5} ($m_c/m_r$ times); and
the same micro-tile $C_r$ is accessed once per iteration of loop \textsf{L6} ($k_c$ times).
Thus, the cost of retrieving the corresponding data block ($B_c, A_c, B_r$, or $C_r$) from one level of the memory
hierarchy is partially (or totally) amortized by re-using it multiple times.
In some detail, 
for example, the micro-tile $C_r$ is loaded from the global memory to the AIE tile vector registers once during the execution 
of the micro-kernel. This requires $m_rn_r$ reads and the same number of writes. In exchange, we perform
$2m_rn_rk_c$ arithmetic operations with that data. Thus, provided $k_c$ is large, the cost of the reads from/writes to
global memory is small and can be hidden/overlapped with the arithmetic.\\

\noindent
\textbf{Copy $A_c,B_c$ from global memory to FPGA.}
Matrices $A$ and $B$ are initially placed in the DDR memory. 
At each iteration of loops \texttt{L2} and \texttt{L3}
a block of $B$ and a block of $A$ are respectively 
packed and copied into the FGPA Block RAM and Ultra RAM.
Due to the high reuse factors, 
the cost of packing these matrices into the corresponding buffers 
is negligible provided the dimensions of \gemm are large. Therefore, in the next section, we omit this cost from the experimental study via emulation.\\

\noindent
\textbf{Copy $B_r$ from FPGA to local memory.}
In reference to the copying of micro-tiles $B_r$, 
we spent a significant effort to obtain a greater computational throughput from the micro-kernel. 
Initially, we employed the GMIO interface to transfer micro-tiles $B_r$ to the local memory of the AIE tiles. 
This method requires two buffers in the local memory to support the GMIO interface. 
Specifically, when GMIO is leveraged to transmit $K$~KB of data to the local memory, 
the compiler will allocate a $K$-KB ping buffer plus a $K$-KB pong buffer in the local memory to implement the protocol. 
Furthermore, once these buffers have been consumed, they cannot be re-purposed for another task at runtime. 
In consequence, the utilization of GMIO for transferring 10~KB of data for the micro-panel $B_r$ 
necessitates an additional 20~KB of memory for buffering,  consuming 30~KB out of the total 32-KB local memory.
This strongly limits the practical dimension of the micro-panel that can be maintained in the local memory
(and the value of the parameter $k_c$).

To tackle this problem, we transitioned to the streaming interface, eliminating the need for buffers in the local memory. With this solution, most of the 32-KB local memory is occupied by 
the micro-panel $B_r$. This allows to use a larger value for $k_c$, augmenting the compute-to-communication ratio for
the micro-kernel: $2 m_r n_r k_r / (2m_rn_r + m_rk_c + n_rk_c)$.
To validate this observation, we experimentally compared the number of MAC operations per cycle for the two kernel designs. 
The first design, which employs buffers and dedicates 8~KB of the AIE tile local memory to the micro-panel $B_r$, delivered 30 MACs/cycle. The second design, which increases the space of the local memory dedicated to accommodating $B_r$, 
in practice allowing a larger value of $k_c$, delivered 37.4 MACs/cycle.\\\

\noindent
\textbf{Stream rows of $A_r$ from FPGA to vector registers.}
In our design of \gemm, with parallelism extracted from loop \textsf{L4},
all AIE tiles access the same micro-panel $A_r$, residing as part of the buffer $A_c$ in the FPGA Ultra RAM,
during the execution of the corresponding micro-kernel.
The selected stream-to-stream multicasting interface has a substantial bandwidth and scalability, making it a good choice.\\

\noindent
\textbf{Transfer $C_r$ from global memory to vector registers.}
In contrast, the smaller and less often accessible micro-tiles of $C_r$ employ slower DDR memory to reduce stress on other interfaces. The data transmission from DDR memory to the AIE tiles was achieved by utilizing the GMIO interface. 
\section{Performance Analysis}\label{sec:profiling}

Section~\ref{sec:parallelGEMM} described our approach to map \gemm to the AMD Versal platform, exploiting 
the multiple AIE tiles by distributing the dimension $n$ (i.e., loop \textsf{L4}) 
of the problem across the individual tiles. 
In this section we assess the effective performance and scalability of the approach, for up to 32~AIE tiles, measuring the
execution time and throughput of the algorithm for a fixed-size problem (strong scaling)
with $(m,n,k)=(m_c,n_c,k_c)=$(256,256,2048).
We select this dimension because the \gemm algorithm decomposes larger problems
into subtasks of size $(m_c,n_c,k_c)$. For large problems, we can expect
the execution time to be linearly proportional to the number of subtasks 
executed, and the performance to be equivalent to that observed when 
running a single subproblem.

\begin{table}
    \centering
    \begin{tabular}{r|rrr|r}

    \hline 
    \#AIE         &  \multicolumn{3}{c|}{Instruction Cycles}      & Performance/tile \\
    tiles         &  {Copy $C_r$} & Arithmetic &  {Total}     & (in MACs/cycle)  \\
    \hline\hline
        1             &  40  & 4,110  &  $3694.1 \cdot 10^{3}$ & 31.5 \\  
        2             &  58  & 4,110  &  $1916.0 \cdot 10^{3}$ & 31.4 \\
        4             &  63  & 4,110  &   ~$958.1 \cdot 10^{3}$ & 31.3 \\
        8             &  84  & 4,110  &   ~$498.9 \cdot 10^{3}$ & 31.2 \\
        16            &  157 & 4,110  &   ~$275.3 \cdot 10^{3}$ & 30.7 \\
        32            &  282 & 4,110  &   ~$162.9 \cdot 10^{3}$ & 29.8 \\       
    \hline
    \end{tabular}
    \caption{Distribution of execution time (in cycles) and performance of the parallel design for \gemm when
    varying the number of AIE tiles between 1 and 32, for a problem of fixed dimension 
    $(m_c,n_c,k_c)=(256,256,2048)$.}
\label{tab:multi_1to32_aie_time_performance}
\end{table}

\subsection{Transfer costs for the micro-kernel}

To gain a better understanding of the execution time, it is crucial to expose the essential costs of the data transfers
that occur inside the micro-kernel. 
For this purpose, 
we assume that the buffers $A_c$ and $B_c$  are already packed in the FPGA Ultra and Block RAMs. 
For a large problem,
the costs of copying and packing $A_c, B_c$ into the corresponding levels of the memory hierarchy are amortized over a significant number of iterations:
$n_c/n_r$ for the former and $m/m_c$ for the latter. 
Therefore we will not consider those transfers during the following elaboration.

At every iteration of loop~\textsf{L4}, 
each AIE tile must first load a distinct micro-panel $B_r$ into its local memory. 
Although this copy is relatively slow (and, therefore, costly), it is amortized 
over all the iterations of loop~\textsf{L5}; that is, $m_c/m_r$ times.
For our particular problem, the cost of this specific transfer from the FPGA Ultra RAM remains constant at 3,280 cycles per copy,
independently of the number of AIE tiles, 
which shows that all AIE tiles perform this transfer simultaneously.

In addition to 
copying $B_r$, each AIE tile has to retrieve a distinct micro-tile $C_r$ from the global
memory, for accumulating the partial contribution of the micro-kernel
execution to the final result. The time cost for this transfer is labeled as 
``Copy $C_r$'' in Table~\ref{tab:multi_1to32_aie_time_performance}, 
and reports 
the time for loading/storing $C_r$ from/to the global memory. 
The results in the table indicate that the time required for this type of copy is negligible when using only one AIE tile
(40 cycles), but consistently grows with the number of AIE tiles (282 cycles for 32~AIE tiles). 
The reason is that increasing the number of AIE tiles
implies the utilization of several GMIO interfaces. 
However, access to the DDR is intrinsically serial, resulting in additional delay when many GMIOs are used. 

The third factor contributing to the transfer costs for the micro-kernel is the time required to 
read the entries of the micro-panel $A_r$, through streaming, from the FPGA Ultra RAM.
The cost associated with this type of streaming is factored in as part of the total execution time measurements 
presented in Table~\ref{tab:multi_1to32_aie_time_performance} (column with label ``Total''). 
In this case, the transmission of $A_r$ benefits from multicasting (since the same rows of $A_r$ are to be read by all
AIE tiles), enabling the data to be received simultaneously.
Eliminating the cost of the arithmetic from the total cost exposes that 
multicasting a single 64-element vector of $A_r$
(to be loaded into either 
\texttt{ar0} or
\texttt{ar1} via the intrinsic
\texttt{readincr\_v64()})
takes approximately 19~cycles, independently of the number of AIE tiles. 

\subsection{Arithmetic cost for the micro-kernel}

Taking into account the unrolling factor applied to loop \textsf{L6},
for the execution of a single micro-kernel with $k_c=2048$, an AIE tile takes $k_c/16$ iterations,
at each invoking 8 times the intrinsic \texttt{mac16()} to compute
$8 \cdot (m_r n_r 16)$ MAC operations; 
that is 1024 MACs with 128 MACs per call to \texttt{mac16()}.
For $k_c=2048$, the total number of MAC operations per micro-kernel is therefore 
$(2048/16) \cdot 1024 = 131072$.
Once the AIE tiles have the necessary data, they can all proceed 
independently (and in parallel) to execute these computations.
The arithmetic performance, without the data transfer costs, is expected to scale linearly with the number
of AIE tiles and proceed at a rate of 128 MACs per cycle.

\subsection{Sustained performance}

The performance profile displayed in \Cref{tab:multi_1to32_aie_time_performance} reports 
the parallel design attains a throughput of 31.5~\macpc using a single AIE tile.
This rate is below the peak performance for a single AIE tile of this platform, which is 128 \macpc for the UINT8 datatype.
This disparity can be attributed to the slow data transfers compared to the high arithmetic throughput. However, confirming this hypothesis requires a deeper analysis of this platform's computation/communication balance.

For this purpose, let us consider a single iteration of the micro-kernel's loop \textsf{L6}. Remember that
the design of the loop body includes 8 calls to the \texttt{mac16()} intrinsic, 
computing a total of 1024 MACs, in principle requiring just 8 cycles for the arithmetic (1 cycle per \texttt{mac16()}).
In comparison, the communication per iteration of the micro-kernel loop body 
involves reading two vectors with 64 elements  each from
$A_r$, with a theoretical cost of 38 cycles (19 cycles per vector). 
(For simplicity, we obviate for the moment the cost of reading the $4 \cdot 32$ elements from $B_r$.)
With these numbers, a rough estimation of performance that can be attained with the micro-kernel is 
given by $1024/38=22.2$ \macpc.
This shows that there is a certain level of internal overlapping between communication and computation
in the design as this theoretical rate is below the experimental performance reported in
\Cref{tab:multi_1to32_aie_time_performance}.

To expose the implicit overlaps, we conducted two experiments:
In the first case we run the micro-kernel with only the transfers of data from $A_r$; and, in the second one,
with the arithmetic operations only.
The results from these experiments are summarized in \Cref{tab:only_ar_mac_experiments} and discussed next.

\begin{table}
    \centering
    \begin{tabular}{l|rr}
    \hline
    Experiments & Measured cycles & Theoretical cycles \\
    \hline\hline
    read ar only         & 4106 &  4864 \\
    execute \texttt{mac16()} only & 1042 & 1024 \\
    baseline             & 4110 &  5888 \\
    \hline
    \end{tabular}
    \caption{Experimental cycle counts and theoretical calculation of the innermost for loop of the micro-kernel when 1) only contain a reading of two micro-tile of $A_r$, 2) only execute \texttt{mac16()} intrinsics. $k_c=2048$. }
    \label{tab:only_ar_mac_experiments}
\end{table}

The first ablated experiment only retains the transfers of vectors \texttt{ar0} and \texttt{ar1} while executing a 
micro-kernel with $k_c=2048$. This setup requires 
the execution of $2048/16=128$ iterations of loop \textsf{L6}, with an expected cost of
$128 \cdot (19+19) =4864$ cycles for the data transfers in the experiment. However, the actual number of cycles is below the theoretical calculation: 
4106 cycles.
We interpret this as an acceleration attained by the compiler/hardware, which rewrites
the read of two 64-element vectors as a single long vector of 128 elements, thus
saving part of the cost.
We validated this hypothesis in a separate experiment.

The second ablated experiment inspects the cost of executing only the arithmetic for a micro-kernel with $k_c=2048$. 
Under these conditions, the micro-kernel execution should take $128 \cdot 8 =1024$ cycles, while the actual measures indicate
a cost of 1042 cycles. The slight variation between these two values can be attributed to the small overheads associated
with the loop control.

In the previous experiments, either our results were consistent with the theoretical costs 
(arithmetic), or we could explain the divergences because
of internal optimizations performed by the compiler/hardware (data transfers for $A_r$).
In these studies though, we analyzed communication and computation separately, resulting in 4106 and 1042 cycles, 
respectively. Now, let us assume no additional optimization occurs when these two components
(arithmetic+data transfers from $A_r$) are combined. The cost should then be 4106 + 1042 $=$ 5148 cycles.
However, the actual experiments show that the cost matches that of reading the elements from $A_r$: 
4,110 cycles.
This indicates a perfect overlap between the arithmetic (and the transfer of $B_r$) and the transfers of $A_r$,
so the total cost is equivalent to that of the heavier component.

The results of this analysis 
show that the actual performance is slightly higher than the theoretical estimation on a single AIE tile: 31.5 \macpc versus 22.2 \macpc.
It also reveals the source of this acceleration: The internal optimizations
performed by the hardware/compiler when reading longer streams of elements 
from $A_r$.
At the same time, it also exposes a vital bottleneck of the platform: The low bandwidth of the FPGA Ultra RAM.
Specifically, in one iteration of loop \textsf{L6} we bring $2 \cdot 64$ UINT8 elements (of $A_r$) from this memory level,
to perform 1024 MAC operations with them. Even though the computation to communication ratio
is $1024/(64 \cdot 2) = 8$ MACs/byte, we clearly see this is not high enough. 
The micro-kernel design is thus limited by the memory bandwidth of the FPGA Ultra RAM design, 
turning it into a communication-bound kernel on this platform. 

\subsection{Scalability of the parallel design}

In this study, we initially established a baseline with a single AIE tile to compute a matrix multiplication 
of dimension $(m,n,k) = (m_c,n_c,k_c) =(256,256,2048)$. 
For this problem, the design required a total time of
\textcolor{black}{$3694.1 \cdot 10^{3}$ cycles},
delivering 31.5 MACs/cycle.
The parallel design is next evaluated under an intense scaling scenario; 
that is, keeping fixed the problem dimensions
while increasing the number of AIE tiles. 
The last column of
Table~\ref{tab:multi_1to32_aie_time_performance} reports the MACs/cycle and
AIE tile when the number of cores is raised from 1 to 32.
These numbers indicate a fair scalability, with the performance per tile decreasing
by a small factor only: from 31.5 MACs/cycle for one AIE tile to
29.8 MACs/cycle for 32 AIE tiles (5.7\%).

\section{Conclusions}

This work investigates the mapping of \gemm to a Versal ACAP using multiple AIE tiles. Matrix multiplication is a well-known, central computational kernel in scientific and engineering applications where 64-bit floating-point precision is paramount.  Here, we instead turn our attention to the deep learning domain and exploit the mixed integer precision on the SIMD units present in the AIE tiles using an architecture-specific microkernel for \gemm. In addition, we mimic the ideas underlying GotoBLAS2 to distribute the matrix operands across the memory hierarchy. For this purpose, since the Versal ACAP lacks a cache memory controller, we use the packing routines and the microkernel itself to orchestrate and execute the data movements. Finally, we parallelize the loop \textsf{L4} of the \gemm algorithm
to match the memory organization of the platform, with a private local memory 
per AIE tile, but a shared FPGA RAM.

We perform a theoretical analysis and a series of experiments to evaluate the performance of the \gemm routine. On the negative side, our results show that the implementation is memory-bound on this platform mostly due to the low bandwidth of the FPGA Ultra RAM. On the positive side, the parallel design is highly scalable, with a parallel efficiency that, in a strong scaling scenario, only degrades by 5\% when increasing the number of AIE tiles from 1 to 32.

\section*{Acknowledgments}
The authors gratefully acknowledge funding from the EuropeanUnion’s Horizon 2020 Research and Innovation Programme under theMarie Skłodowska Curie grant agreement No. 956090 (APROPOS, http://www.apropos-itn.eu/).

This work also received funding in Spain from the 
research project
PID2020-113656RB-C22 of MCIN/AEI/10.13039/501100011033,
y por FEDER \textit{Una manera de hacer Europa}.
as well as from 
European High-Performance Computing Joint Undertaking (JU) under grant agreement No.
955558 (eFlows4HPC project). The JU receives support from the European Union’s Horizon 2020 research and innovation program,
and Spain, Germany, France, Italy, Poland, Switzerland, Norway.

\bibliographystyle{unsrt}  
\bibliography{references, E_BIB}

\end{document}